\documentclass[aps,preprint,amsmath,amssymb]{revtex4}
\usepackage{graphicx}% Include figure files
\def\be{\begin{eqnarray}}
\def\en{\end{eqnarray}}
\def\non{\nonumber}

\def\pl{{ Phys. Lett.}~}
\def\pr{{ Phys. Rev.}~}

\def\bi{\bibitem}
\begin{document}

\title{\Large \bf Combined Chiral Dynamics and MIT Bag Model Study of Strong
 $\Sigma^{(*)}_Q \to \Lambda_Q\pi$ Decays
 %Strong $\Sigma_b \to \Lambda_b\pi$ Decay in MIT bag Model
 }

\author{ \bf \large Chien-Wen Hwang\footnote{Email: t2732@nknucc.nknu.edu.tw}
 }

\affiliation{ Department of Physics, National Kaohsiung Normal University, Kaohsiung, Taiwan 824,
Republic of China
 }

%\date{\today}

\begin{abstract}
The strong decays of the heavy baryons $\Sigma^{(*)}_Q \to \Lambda_Q \pi$ are studied by combining
the chiral dynamics and the MIT bag model. In charm sector, we calculate the decay widths
$\Gamma(\Sigma^{(*)}_c\to \Lambda_c\pi)$ and compare with the experimental data and other theoretical
estimations.  In addition, we also predict the strong decay widths $\Gamma(\Sigma^{(*)}_b\to
\Lambda_b\pi)$.
\end{abstract}

\maketitle %
%%%%%%%%%%%%%%%%%%%%%%%%%%%%%%%%%%%%%%%%%%%%%%%%%%%%%%%%%%%%%%%%%%%%%%%

\section{Introduction}
Two exotic relatives of the proton and neutron $\Sigma^{(*)+}_b$ ($buu$) and $\Sigma^{(*)-}_b$
($bdd$) were discovered by CDF collaboration at Fermilab \cite{CDF}. Their masses are:
$M_{\Sigma^+_b}=5808^{+2.0}_{-2.3} \pm1.7$ MeV, $M_{\Sigma^-_b}=5816^{+1.0}_{-1.0} \pm1.7$ MeV,
$M_{\Sigma^{*+}_b}=5829^{+1.6}_{-1.8} \pm1.7$ MeV, and $M_{\Sigma^{*-}_b}=5837^{+2.1}_{-1.9} \pm1.7$
MeV. Due to the mass of bottomed baryon $\Lambda^0_b$ is $5624 \pm 9$ MeV \cite{PDG06}, and the mass
differences $M_{\Sigma^{(*)}_b}-M_{\Lambda_b}$ are all larger than $M_\pi$, thus the dominate decay
modes of these exotic bottomed baryons are the strong decays $\Sigma_b^{(*)} \to \Lambda_b \pi$,
which are similar to the ones of charmed baryons $\Sigma_c^{(*)} \to \Lambda_c \pi$. In this paper,
we will combine the chiral dynamics and MIT bag model to study the strong decays of the heavy baryon
$\Sigma^{(*)}_Q \to \Lambda_Q \pi$.

The light quark contribution to the QCD Lagrangian
 \be
 {\cal L}_q = {\bar q}(i\not\!\! D-m_q)q,~~~~(q=u, d, {\textrm{and}}~s)
 \en
has an approximate SU(3)$_L$ $\times$ SU(3)$_R$ flavor chiral symmetry, because the current quark
masses are all very small on the typical hadron energy scale. It was known early on that this
symmetry must be spontaneously broken by the QCD vacuum, so that the physical spectra of hadrons made
up of light quarks have only SU(3)$_{L+R}$ symmetry. Moreover, due to the spontaneous breaking of the
chiral symmetry, there exist eight pseudoscalar bosons (called Goldstone bosons, which include three
$\pi$'s, four $K$'s, and one $\eta$). Their couplings to hadrons at low energies are determined by
partial conservation of axial-vector current (PCAC) and current algebra. The chiral properties of a
heavy hadron is dictated by its light quark contents. For the heavy baryon, a heavy quark $Q$ will
combine the two light quarks to form baryons $Qq_1q_2$. However, the two light quarks can form a
symmetric sextet or an antisymmetric antitriplet in flavor SU(3) space. For the ground state baryons,
the SU(3) symmetric sextet diquarks have spin 1 whereas the SU(3) antisymmetric antitriplet diquarks
have spin 0. Thus for the SU(3) symmetric sextet there are both spin $\frac{1}{2}$ baryons ($B_6$)
and spin $\frac{3}{2}$ baryons ($B^*_6$). For the SU(3) antisymmetric antitriplet there is only spin
$\frac{1}{2}$ baryons ($B_{\bar 3}$). Once the flavor SU(3) contents of these heavy baryons are
determined, their couplings to Goldstone bosons can be immediately written down following the
standard formalism of chiral dynamics.

The MIT bag model \cite{CJJ,CJJ1} is a simple relativistic model of hadrons which is consistent with
some of the essential features of QCD, namely confinement, gluon degree of freedom, and gauge
invariance. It has been applied to describe hadron spectroscopy \cite{DJJK,ID} and various
transitions where momentum transfers are not large \cite{ID}. The greatest success of the MIT bag
model lies in its description of ground state hadrons formed with light quarks, where agreement with
the experimental data is remarkable.

The paper is organized as follows. In Sec. II we review the basic MIT bag model formalism and
construct the heavy baryon wave functions. In Sec. III  we consider the dynamics of heavy baryons
interacting with the Goldstone bosons. We discuss the chiral properties of the heavy baryons and
derive the chiral Lagrangian involving heavy baryons. Applications and numerical results for the
strong decay widths of some heavy baryons are worked out in Sec. IV. Finally, the conclusion is given
in Sec. V.
%%%%%%%%%%%%%%%%%%%%%%%%%%%%%%%%%%%%%%%%%%%%%%%%%%%%%%%%%%%%%%%%%%%%%%%%%

\section{Formalism of MIT bag model for heavy baryon}
\def\uds%
{\begin{array}{c}
u \\
d \\
s
\end{array}}
\def\bsix%
 {\begin{array}{ccc}
 {\Sigma_Q^{+1}}~  & {1\over{\sqrt {2}}}\Sigma^0_Q & ~{1\over{\sqrt {2}}}\Xi'^{+1/2}_Q \\
 {1\over{\sqrt {2}}}\Sigma^0_Q & {\Sigma_Q^{-1}} & {1\over{\sqrt {2}}}\Xi'^{-1/2}_Q \\
 {1\over{\sqrt {2}}}\Xi'^{+1/2}_Q & {1\over{\sqrt {2}}}\Xi'^{-1/2}_Q  & \Omega_Q
\end{array}}
\def\bthree%
{\begin{array}{ccc}
 0~  & ~~~~~~\Lambda_Q~~~~~~~ & ~\Xi^{+1/2}_Q \\
 -\Lambda_Q & 0 & \Xi^{-1/2}_Q \\
 -\Xi^{+1/2}_Q & \Xi^{-1/2}_Q  & 0
\end{array}}
\def\bsixs%
 {\begin{array}{ccc}
 {\Sigma_Q^{*+1}}~  & {1\over{\sqrt {2}}}\Sigma^{*0}_Q & ~{1\over{\sqrt {2}}}\Xi'^{+1/2}_Q \\
 {1\over{\sqrt {2}}}\Sigma^{*0}_Q & {\Sigma_Q^{*-1}} & {1\over{\sqrt {2}}}\Xi'^{*-1/2}_Q \\
 {1\over{\sqrt {2}}}\Xi'^{*+1/2}_Q & {1\over{\sqrt {2}}}\Xi'^{*-1/2}_Q  & \Omega^*_Q
\end{array}}

The MIT bag model is essentially a relativistic shell model which describes quarks moving
independently inside a confining spherical cavity of radius $R$. The bag Lagrangian for quark only is
\begin{equation}
     {\cal L}_{bag} = \left\{{i\over {2}} [\bar \psi \gamma^\mu \partial_\mu \psi-(\partial_\mu \bar \psi)
     \gamma^\mu \psi] - m\bar \psi \psi - {\cal B}\right\}\theta_V (r) - {1\over {2}}\bar \psi \psi \Delta_s (r),                                     \label{Lag_f}
\end{equation}
\noindent where $m$ is the quark mass, $\theta_V (r) = \theta (R-r), \Delta_s (r) = \delta (R-r)$,
the constant ${\cal B}$ is the volume energy density, which comes from the work done against the QCD
vacuum in creating the cavity, finally the surface term is added to ${\cal L}_{bag}$ so that the
quarks move as if they had an infinite mass outside the bag \cite{Bog}.

The Euler-Lagrange equations of motion are
 \be
 &&(i\gamma^\mu \partial_\mu - m) \psi = 0~~~~~~~~~~~~~~~~{\rm inside~the~bag}, \label{eom1} \\
 &&\begin{array}{c}
       i\gamma^\mu n_\mu \psi = \psi \\
 n_\mu \cdot \partial ^\mu (\bar \psi \psi) = 2 {\cal B}
   \end{array} ~~~\Bigg\}~~~~~~~~~~{\rm
 on~the~bag~surface}, \label {eom2}
 \en
where $n_\mu$ is a covariant inward vector normal to the bag surface, $n_\mu = (0,-\hat r)$ for a
static spherical bag. From Eq.~(\ref{eom1}), we see that the Lagrangian density Eq.~(\ref{Lag_f})
yields a free Dirac equation inside the bag as expected. The quadratic boundary condition (the second
line of Eq.~(\ref{eom2})) requires that only $j=1/2$ modes can be excited within a static rigid bag.
Consequently only $S$ and $P$ $(l=0,1)$ orbital angular momenta are allowed, and the single particle
excitation spectrum consists of two classes of states with total angular momentum ${j=1/2}$, namely
$n S_{1/2}$ and $n P_{1/2}$ ($n=$ radial quantum number). Explicit expressions for these solutions
are \cite{ID}
 \be u^0_n = N^0_n\left(\begin{array}{c}
            \sqrt{({\omega + m\over {\omega}})}~ij_0(X_n {r\over {R}})~\chi \\
            - \sqrt{({\omega - m\over {\omega}})}~j_1(X_n {r\over {R}}) \sigma \cdot \widehat{r}~\chi
                   \end{array} \right) ~e^{-i\omega t},    \label{sol0}
 \en
and
 \be
 u^1_n = N^1_n\left(\begin{array}{c}
            \sqrt{({\omega + m\over {\omega}})}~ij_1(X_n {r\over {R}}) \sigma \cdot \widehat{r}~\chi \\
             \sqrt{({\omega - m\over {\omega}})}~j_0(X_n {r\over {R}})~\chi       \end{array} \right)~e^{-i\omega t} ,                              \label{sol1}
 \en
where the normalization coefficients $N^0_n, N^1_n$ are fixed by the integral $\int_R \psi^{\dag}
\psi d^3 r = 1$, $j_0, j_1$ are spherical Bessel functions
 \be
 j_0(x)={\sin x\over{x}},~~~j_1(x)={\sin x\over{x^2}}-{\cos x\over{x}},
 \en
$\chi$ is a Pauli spinor, $X_n$ is the quark's momentum times $R$, and $\omega$ is the eigenenergy
given by
 \be
     \omega = \sqrt{(m^2+{X_n^2\over R^2})}.       \label{energy}
 \en

The eigenvalue equations can be derived by substituting Eqs.~(\ref{sol0}) and (\ref{sol1}) into the
first line of Eq.~(\ref{eom2}). For $u^0_n$, we get
 \be \left(\begin{array}{cc}
 0 & -i \sigma \cdot \widehat{r} \\
 i \sigma \cdot \widehat{r} & 0  \\
 \end{array}\right)
 \left(\begin{array}{c}
            \sqrt{({\omega + m\over {\omega}})}~ij_0(X_n) \\
            - \sqrt{({\omega - m\over {\omega}})}~j_1(X_n) \sigma \cdot \widehat{r}   \end{array} \right)
 =\left(\begin{array}{c}
            \sqrt{({\omega + m\over {\omega}})}~ij_0(X_n) \\
            - \sqrt{({\omega - m\over {\omega}})}~j_1(X_n) \sigma \cdot \widehat{r}   \end{array} \right),
            \label{constraint1}
 \en
which leads to
 \be
   j_1(X_n)=\sqrt {({\omega+m\over {\omega-m}})}~j_0(X_n),  \label{constraint2}
 \en
and finally
 \be
    \tan X_n = {X_n\over {1-m R-\sqrt {m^2R^2 + X_n^2}}}. \label{212}
 \en
The corresponding eigenvalue equation for the $u_n^1$ mode is
 \be
    \tan X_n = {X_n\over {1-m R+\sqrt {m^2R^2 + X_n^2}}}.  \label{213}
 \en
When $m \rightarrow 0$, Eqs.~(\ref{212}) and (\ref{213}) yield the transcendental equations
 \be
     \tan X_n = {X_n\over {1 \mp X_n}}. \label{312}
 \en
 Some of the low lying solutions to Eq.~(\ref{312}) are listed in TABLE I \cite{Ba}.
%\vskip 1cm
\begin{table}[htb]
\begin{center}
\begin{tabular}{|c||c|c|c|c|c|c|} \hline
 ~state~ & $1S_{1/2}$ & $1P_{1/2}$ & $2S_{1/2}$ & $2P_{1/2}$ & $3S_{1/2}$ & $3P_{1/2}$
  \\ \hline
 $X_n$ & 2.043 & 3.812 & 5.396 & 7.002 & 8.578 & 10.163   \\ \hline
\end{tabular}
\end{center}
%\begin{center}
\caption{Bag model eigenmodes for a massless quark.}
\end{table}
%\vskip 1.5cm

After solving the bag equations of motion inside a cavity, we can expand a second quantized light
quark field operator $\psi_q (x)$ in terms of the quark eigenmodes,
 \be
   \psi_q (x) = \sum_{nlm} b_{nlm} u_{nlm} (\vec r,t) + d^\dagger_{nlm} v_{nlm} (\vec r,t)
 \en where $b_{nlm}$ is the canonical quark annihilation operator and $d^\dagger_{nlm}$ is the
antiquark creation operator. These operators satisfy the usual anticommutation relations, namely
 \be
   \left\{ b_{nlm},b^\dagger_{n'l'm'} \right\}=\left\{ d_{nlm},d^\dagger_{n'l'm'} \right\} =
   \delta_{nn'}\delta_{ll'}\delta_{mm'},
 \en
with the others equal to zero. We can then readily construct a state of heavy baryon with spin $S$ as
 \be
   |B_Q^{nl} \rangle = C(S,S_z,s_1,s_2,s_3) |q^{nl}_1(s_1) q^{nl}_2 (s_2) Q(s_3)\rangle
 \en
where $C(S,S_z,s_1,s_2,s_3)$ is the familiar Clebsch-Gordan coefficient. Each light quark is in a
triplet
 \be
 q=\left(
   \uds
   \right)
 \en
of the flavor SU(3). Since $3\otimes 3 = 6\oplus\bar 3$ and the lowest lying light quark state has
$n=1$ and $l=0$ ($S$-wave), there are two different diquarks: a symmetric sextet ($s_1+s_2=1$) and an
antisymmetric antitriplet ($s_1+s_2=0$). When the diquark combines with a heavy quark, the sextet
contains both spin $\frac{1}{2}$ ($B_6$) and  spin $\frac{3}{2}$ ($B^*_6$) baryons, and the
antitriplet contains only spin $\frac{1}{2}$ ($B_{\bar 3}$) baryons. Explicitly, the wave functions
of spin $\frac{1}{2}$ $B_6$ baryon are
 \be
 &&|\Sigma^{+1}_Q \uparrow\rangle = \sqrt{2\over{3}}|uu\rangle|\uparrow\uparrow\rangle|Q\downarrow\rangle
 -\sqrt{1\over{3}}|uu \rangle\sqrt{1\over{2}}(|\uparrow\downarrow\rangle+|\downarrow\uparrow\rangle)|Q\uparrow\rangle, \non \\
 &&|\Sigma^{0}_Q \uparrow\rangle = \sqrt{2\over{3}}\sqrt{1\over{2}}(|ud\rangle+|du\rangle)|\uparrow\uparrow\rangle|Q\downarrow\rangle
 -\sqrt{1\over{3}}\sqrt{1\over{2}}(|ud\rangle+|du\rangle)\sqrt{1\over{2}}(|\uparrow\downarrow\rangle+|\downarrow\uparrow\rangle)|Q\uparrow\rangle, \non \\
 &&|\Sigma^{-1}_Q \uparrow\rangle = |\Sigma^{+1}_Q \uparrow\rangle_{(u \to d)},~~~~~~~~~~~~~~~
 |\Xi'^{+1/2}_Q \uparrow\rangle = |\Sigma^{0}_Q \uparrow\rangle_{(d \to s)}, \non \\
 &&|\Xi'^{-1/2}_Q \uparrow\rangle = |\Xi'^{+1/2}_Q \uparrow\rangle_{(u \to d)},~~~~~~~~~~
 |\Omega_Q\uparrow\rangle = |\Sigma^{+1}_Q \uparrow\rangle_{(u \to s)}, \label{b66}
 \en
where the superscript denotes the value of the isospin quantum number $I_3$. An asterisk on the
symbol will denote a corresponding spin $\frac{2}{3}$ baryon:
 \be
 &&|\Sigma^{*+1}_Q \frac{1}{2}\rangle = \sqrt{1\over{3}}|uu\rangle|\uparrow\uparrow\rangle|Q\downarrow\rangle
 +\sqrt{2\over{3}}|uu \rangle\sqrt{1\over{2}}(|\uparrow\downarrow\rangle+|\downarrow\uparrow\rangle)|Q\uparrow\rangle, \non \\
 &&|\Sigma^{*0}_Q \frac{1}{2}\rangle = \sqrt{1\over{3}}\sqrt{1\over{2}}(|ud\rangle+|du\rangle)|\uparrow\uparrow\rangle|Q\downarrow\rangle
 +\sqrt{2\over{3}}\sqrt{1\over{2}}(|ud\rangle+|du\rangle)\sqrt{1\over{2}}(|\uparrow\downarrow\rangle+|\downarrow\uparrow\rangle)|Q\uparrow\rangle, \non \\
 &&|\Sigma^{*-1}_Q \frac{1}{2}\rangle = |\Sigma^{+1}_Q \uparrow\rangle_{(u \to d)},~~~~~~~~~~~~~~~
 |\Xi'^{*+1/2}_Q \frac{1}{2}\rangle = |\Sigma^{0}_Q \uparrow\rangle_{(d \to s)}, \non \\
 &&|\Xi'^{*-1/2}_Q \frac{1}{2}\rangle = |\Xi'^{*+1/2}_Q \frac{1}{2}\rangle_{(u \to d)},~~~~~~~~~~
 |\Omega^*_Q\frac{1}{2}\rangle = |\Sigma^{*+1}_Q \frac{1}{2}\rangle_{(u \to s)}, \label{b66ss}
 \en
As to the wave functions of spin $\frac{1}{2}$ $B_{\bar 3}$ baryon, they are
 \be
 &&|\Lambda_Q \uparrow\rangle =\sqrt{1\over{2}}(|ud\rangle-|du\rangle)\sqrt{1\over{2}}
 (|\uparrow\downarrow\rangle-|\downarrow\uparrow\rangle)|Q\uparrow\rangle, \non \\
 &&|\Xi^{+1/2}_Q \uparrow\rangle = |\Lambda_Q \uparrow\rangle_{(d \to s)},~~~~~~~~
 |\Xi^{-1/2}_Q \uparrow\rangle = |\Xi^{+1/2}_Q \uparrow\rangle_{(u \to d)}. \label{b33}
 \en
We may use the decomposition
 \be
 q_{1i}q_{2j}&=&\frac{1}{2}(q_{1i}q_{2j}+q_{1j}q_{2i})+\frac{1}{2}(q_{1i}q_{2j}-q_{1j}q_{2i}) \non \\
 &=& (B_6)_{ij}+ \frac{1}{\sqrt 2}(B_{\bar 3})_{ij}
 \en
to assemble the sextet and the antitriplet in a systematic and an antisymmetric $3\times3$ matrix,
respectively,
 \be
  B_6 &=& \left[
     \bsix
     \right],  \label{B66} \\
 B_{\bar 3} &=& \left[
    \bthree
     \right], \label{B33}
 \en
and a matrix for $B^*_6$
 \be
  B^*_6 &=& \left[
     \bsixs
     \right].  \label{B66ss}
 \en
%%%%%%%%%%%%%%%%%%%%%%%%%%%%%%%%%%%%%%%%%%%%%%%%%%%%%%%%%%%%%%%%%
\section{Chiral Dynamics of Heavy Baryons}
\def\pke%
{\begin{array}{ccc}
{\pi^0 \over {\sqrt{2}}} + {\eta \over {\sqrt{6}}} & \pi^+ & K^+ \\
\pi^- & -{\pi^0 \over {\sqrt{2}}} + {\eta \over {\sqrt{6}}} & K^0 \\
K^- & \bar K^0 & -\sqrt {2 \over {3}} \eta
\end{array}}

Before discussing the interaction between heavy baryons and Goldstone bosons, we will first summarize
the case of Goldstone bosons interacting among themselves \cite{HYC1,Geob}. The chiral symmetry is
nonlinearly realized by using the unitary matrix
 \be
   \Sigma = e^{2iM/\sqrt{2}f_\pi}
 \en
where $M$ is a $3 \times 3$ matrix for the octet of Goldstone bosons
 \be
  M= \left[
     \pke
     \right] \label{uds}
 \en
and $f_\pi = 93$ MeV is the pion decay constant. Under $SU(3)_L \times SU(3)_R$, $\Sigma$ transforms
as
 \be
   \Sigma \to \Sigma' = L \Sigma R^\dagger.
 \en
Hence the lowest order Lagrangian for the Goldstone bosons is
 \be
   {\cal L}_M = {f^2_\pi \over {4}}~Tr~\partial _\mu \Sigma^\dagger \partial^\mu \Sigma.
 \en
In order to facilitate the discussions of the Goldstone bosons interacting with heavy mesons, we
introduce a new Goldstone-boson matrix $\xi \equiv \Sigma^{1/2}$, which transforms under an $SU(3)_L
\times SU(3)_R$ as
 \be
   \xi \to \xi' = L~\xi~U^\dagger = U~\xi~R^\dagger,
 \en where $U$ is an unitary matrix depending on $L$, $R$, and $M$, so that it is no longer global.
Now with the aid of $\xi$, we construct a vector field ${V}_\mu$ and an axial vector field ${A}_\mu$:
 \be
   {V}_\mu &=& {1\over {2}}(\xi^\dagger \partial_\mu \xi + \xi \partial_\mu \xi^\dagger), \label{vector}\\
   {A}_\mu &=& {i\over {2}}(\xi^\dagger \partial_\mu \xi - \xi \partial_\mu \xi^\dagger), \label{axial}
 \en
with the simple transformation properties
 \be
   {V}_\mu \to {V}'_\mu &=& U~{V}_\mu~U^\dagger + U~\partial_\mu~U^\dagger, \label{Vec}\\
   {A}_\mu \to {A}'_\mu &=& U~{A}_\mu~U^\dagger.
 \en
The chiral transformation of $q$ is given in Eq.~(\ref{uds}) which is however not convenient for our
purpose. By means of the following field redefinition \cite{Cheungz}
 \be
   q_L \to \xi ^\dagger q_L,~~~q_R \to \xi q_R,
 \en
the light quarks can be made to transform simply as
 \be
   q \to q' = U~q.  \label{qU}
 \en
and the chiral transformations of the heavy baryons in Eqs.~(\ref{B66}) and (\ref{B33}) can be
established as
 \be
 B_6\to B'_6=UB_6U^T, \\
 B_{\bar 3}\to B'_{\bar 3}=UB_{\bar 3}U^T.
 \en
and the covariant derivatives under chiral transformations for $B_6$ and $B_{\bar 3}$ are
 \be
 D_\mu B_6\equiv \partial_\mu B_6+V_\mu B_6+B_6V^T_\mu, \\
 D_\mu B_{\bar 3}\equiv \partial_\mu B_{\bar 3}+V_\mu B_{\bar 3}+B_{\bar 3}V^T_\mu.
 \en
A similar equations hold for $B^*_6$ and $D_\mu B^*_6$. Then the chiral-invariant Lagrangian is
 \be
 {\cal L}_B&=&{\rm tr} [\bar B_{6}(i\not\!\!D-M_{6})B_{6}]+\frac{1}{2} {\rm tr} [\bar B_{\bar 3}
 (i\not\!\!D-M_{\bar 3})B_{\bar 3}] \non \\
 &+&{\rm tr} \{\bar B^{*\mu}_{6}[-g_{\mu\nu}(i\not\!\!D-M^*_{6})+i(\gamma_\mu D_\nu+\gamma_\nu D_\mu)-
 \gamma_\mu(i\not\!\!D+M^*_{6})\gamma_\nu]B^{*\nu}_6\} \non \\
 &+& g_1 {\rm tr}(\bar B_{6}\gamma_\mu\gamma_5 A^\mu B_{6})+g_2 {\rm tr}(\bar B_{6}\gamma_\mu\gamma_5 A^\mu B_{\bar
 3})+{\rm H.c.} \non \\
 &+& g_3 {\rm tr}(\bar B^*_{6\mu} A^\mu B_{6})+{\rm H.c.}+g_4 {\rm tr}(\bar B^*_{6\mu} A^\mu B_{\bar
 3})+{\rm H.c.} \non \\
 &+& g_5 {\rm tr}(\bar B^*_{6 \nu}\gamma_\mu\gamma_5 A^\mu B^{*\nu}_{6})+g_6 {\rm tr}(\bar B_{\bar 3}
 \gamma_\mu\gamma_5 A^\mu B_{\bar 3}), \label{Lagrangian}
 \en
where $B^*_{6\mu}$ is a Rarita-Schwinger vector-spinor field a spin ${2\over {3}}$ particle \cite{RS}
and $A_\mu$ is the axial field introduced in Eq.~(\ref{axial}).

%%%%%%%%%%%%%%%%%%%%%%%%%%%%%%%%%%%%%%%%%%%%%%%%%%%%%%%%%%%%%%%%%%%%%%%%%%
\section{Applications}
Using the bag model wave functions constructed in the previous section, one can proceed to calculate
the strong decay coupling constants ($g_1 \sim g_6$). However, the six coupling constants can be
reduced to two independent ones by the heavy quark symmetry (HQS) \cite{HYC1}. Furthermore, these two
coupling constants are independent of the heavy mass.

Now we will begin with the model calculation for the coupling constants. The strong coupling
constants describe soft pion emission. Applying PCAC, we can express a soft pion amplitude to a
matrix element of the axial-vector current
 \be
   \langle B' \pi^a (q)|B\rangle = {q^\mu \over {f_\pi}} \langle B' |A^a_\mu |B\rangle.   \label{PCAC}
 \en
Here we consider the strong decay $\Sigma^{+1} _Q \to \Lambda_Q \pi^+ $ firstly. The Lagrangian
Eq.~(\ref{Lagrangian}) leads to the coupling
 \be
 {\cal L}_{\Sigma_Q \Lambda_Q \pi}={g_2\over{\sqrt{2}f_\pi}}{\bar \Sigma}^{+1}_Q\gamma^\mu \gamma_5
 \Lambda_Q\partial_\mu \pi^+. \label{Lpiece}
 \en
If we define $g_A^{\Sigma_Q \Lambda_Q}$ by
 \be
 \langle\Lambda_Q|A^1_\mu -i A^2_\mu|\Sigma_Q^{+1}\rangle= g_A^{\Sigma_Q \Lambda_Q}(q^2)
 \bar u (\Lambda_Q)\gamma_\mu \gamma_5 u(\Sigma^{+1}_Q)+\cdots, \label{gSL}
 \en
where the unlisted terms vanish at $q=0$. Combined Eqs. (\ref{PCAC}) and (\ref{gSL}) and compared
with Eq. (\ref{Lpiece}), we obtain
 \be
 g_2=-g_A^{\Sigma_Q \Lambda_Q}(0)
 \en
The left hand side of Eq. (\ref{gSL}) can be calculated in the MIT bag model, which is
diagrammatically illustrated in FIG. 1.
\begin{figure}[h]
\hskip 8.5cm
~~~~\includegraphics[%
  scale=0.7,
  angle=0]{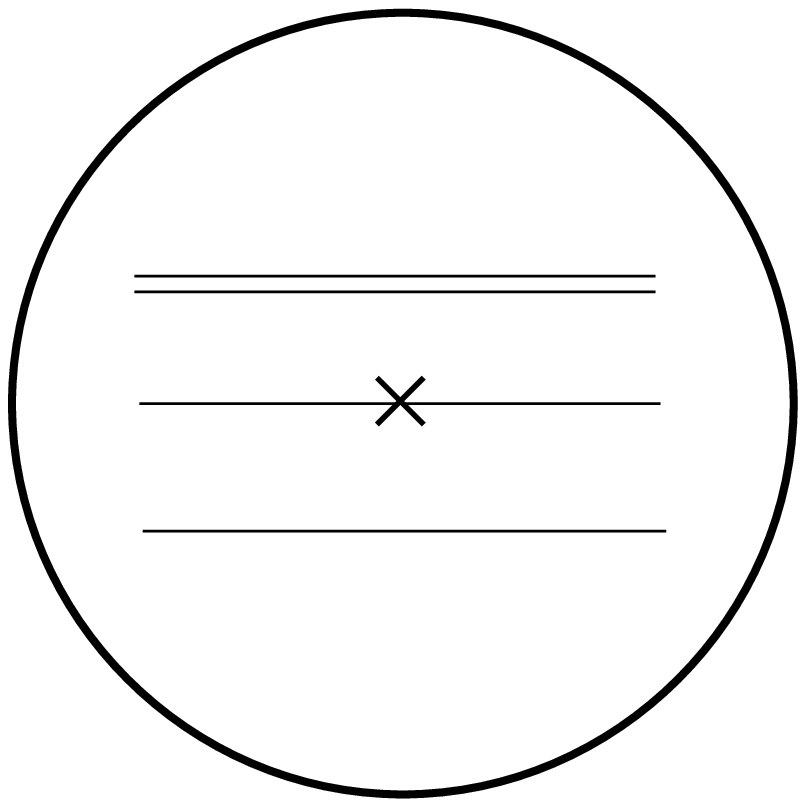}
\caption{Single particle transition in the bag. The single (double) line stands for a light (heavy)
quark, and the cross represents an external current operator.}
\end{figure}
The $\Delta I_3=-1$ transition can be described by the current $\widetilde{\Sigma}_3$ (choosing
$\mu=3$)
 \be
 \widetilde{\Sigma}_3=\int d^3 r (A_3^1-i A_3^2)=\int d^3r~ b^{\dag d}_{nlm} \bar {u}_{nlm}(r,t)
 \gamma_3 \gamma_5 u_{nlm}(r,t) b^u_{nlm},
 \en
where the superscript $u(d)$ represents $u (d)$ quark. Here the light quark states have $n=1$, $l=0$,
and $m=0$ ($S$-wave) because we deal with the $B_6$, $B_6^*$, and $B_{\bar 3}$ baryons. Therefore we
obtain
 \be
  %\langle \Lambda_Q \uparrow |\tilde{\Sigma}_3 |\Sigma^{+1}_Q \uparrow \rangle =
  g_2=
  - {\cal I} \langle \Lambda_Q \uparrow |\widetilde{\Sigma}_3|\Sigma_Q^{+1}\uparrow\rangle,
  \label{amp1}
 \en
where
 \be
 {\cal I}\equiv 4 \pi N_I N_F \int^R_0 {d r}~r^2  \left[ j_0(X_F~\bar r) j_0(X_I~\bar r) - {1\over{3}}
  j_1(X_F~\bar r)j_1(X_I~\bar r)\right], \label{I}
 \en
$\bar r \equiv r/R$, and we have used the identities
 \be
   \vec \sigma \cdot \widehat{r}~\sigma_i~\vec \sigma \cdot \widehat{r} &=& 2~r_i \sigma
   \cdot \widehat{r}- \sigma_i, \non \\
  \int d^3 r f(r) \widehat{r} (\vec \sigma \cdot \widehat{r}) &=& {1\over {3}}\int d^3 r f(r) \vec{\sigma}.
  \non
 \en
Eqs. (\ref{b66}) and (\ref{b33}) then give
 \be
 \langle \Lambda_Q \uparrow |\widetilde{\Sigma}_3|\Sigma_Q^{+1}\uparrow\rangle =
 -\frac{1}{2}\sqrt{\frac{1}{6}}(-4)\langle d \uparrow |\widetilde{\Sigma}_3|u\uparrow\rangle
 =\sqrt{\frac{2}{3}}
 \en
The decay width of the decay $\Sigma^{+1}_Q \to \Lambda_Q \pi$ is
 \be
 \Gamma(\Sigma_Q \to \Lambda_Q \pi)={|\vec{p}|\over{8\pi M^2_{\Sigma_Q}}}g^2_{\Sigma_Q \Lambda_Q
 \pi}[(M_{\Sigma_Q}-M_{\Lambda_Q})^2-M_\pi^2] \label{width1}
 \en
where $\vec p$ is the pion momentum in the c.m. system and
 \be
 g_{\Sigma_Q \Lambda_Q \pi}={M_{\Sigma_Q}+M_{\Lambda_Q}\over{\sqrt{2}f_\pi}}g_2
 \en
is the Goldberger-Treiman relation.

Next, we consider the strong decay $\Sigma^{*+1}_Q \to \Lambda_Q \pi$. The Lagrangian
Eq.~(\ref{Lagrangian}) leads to the coupling
 \be
 {\cal L}_{\Sigma^*_Q \Lambda_Q \pi}={g_4\over{\sqrt{2}f_\pi}}{\bar \Sigma}^{*+1}_{Q\mu}
 \Lambda_Q\partial^\mu \pi^+. \label{Lpiece2}
 \en
We may take the advantage of the constraints imposed by HQS \cite{HYC1}
 \be
 g_4=-\sqrt{3}g_2
 \en
The decay width of the decay $\Sigma^{*+1}_Q \to \Lambda_Q \pi$ is
 \be
 \Gamma(\Sigma^*_Q \to \Lambda_Q \pi)={|\vec{p}|^3\over{96\pi M^2_{\Sigma^*_Q}M^2_{\Lambda_Q}}}
 g^2_{\Sigma^*_Q \Lambda_Q \pi}[(M_{\Sigma^*_Q}+M_{\Lambda_Q})^2-M_\pi^2], \label{width2}
 \en
where
 \be
 g_{\Sigma^*_Q \Lambda_Q \pi}={M_{\Sigma^*_Q}+M_{\Lambda_Q}\over{\sqrt{2}f_\pi}}g_4
 \en
is also the Goldberger-Treiman relation.

The parameters we use as input to calculate ${\cal I}$ are the light quark masses $m_{u,d}$ and the
bag of radius $R$. The current light quark masses in general are taken as $m_{u,d} = 0$. As to the
radius $R$, $R_{\Sigma^{++}_c}=0.945$ fm, $R_{\Sigma^{*++}_c}=1.01$ fm, and $R_{\Lambda^+_c}=0.914$
fm were given by Ref. \cite{BS} in charm sector, and $R_{\Sigma^{+}_b}=1.02$ fm,
$R_{\Sigma^{*+}_b}=1.04$ fm, and $R_{\Lambda^0_b}=0.996$ fm were given by Ref. \cite{Ponce} in bottom
sector. They were obtained by fitting the mass spectrum in the respective sectors. Here we use the
parameter $m_{u,d} = 0$ and $R=1$ fm to Eq. (\ref{I}) and obtain ${\cal I}=0.653$. Then the strong
decay widths in Eqs. (\ref{width1}) and (\ref{width2}) can be estimated, and the numerical results in
charm sector are listed in TABLE II and TABLE III, respectively.
% \vskip 1cm
\begin{table}[htb]
\begin{center}
\begin{tabular}{c|c|c|c} \hline \hline
 & $\Gamma(\Sigma^{++}_c\to\Lambda^+_c \pi^+)$ & $\Gamma(\Sigma^{+}_c\to\Lambda^+_c \pi^0)$ &
 $\Gamma(\Sigma^{0}_c\to\Lambda^+_c \pi^-)$
  \\ \hline
 experiment \cite{PDG06} & $2.23\pm 0.30$ & $< 4.6 (\text {CL} =90\%)$ & $2.2\pm 0.4$    \\
 \textbf{this work} & 1.90 & 2.20 & 1.87\\
 CQM  \cite{CQM1}& $1.31\pm 0.04$ & $1.31\pm 0.04$ & $1.31\pm 0.04$ \\
 CQM  \cite{CQM2}& $2.025^{+1.134}_{-0.987}$ & & $1.939^{+1.114}_{-0.954}$ \\
 HHCPT \cite{HYC1}& 2.47, 4.38 & 2.85, 5.06 & 2.45, 4.35 \\
 HHCPT \cite{HHCPT2}& 2.5 & 3.2 & 2.4 \\
 HHCPT \cite{HHCPT3}& & & $1.94 \pm 0.57$ \\
 HHCPT \cite{HHCPT4}& input & $2.6\pm 0.4$ & $2.2 \pm 0.3$ \\
 LFQM \cite{LFQM}& 1.64 & 1.70 & 1.57 \\
 RTQM \cite{RTQM}& $2.85 \pm 0.19$ & $3.63 \pm 0.27$ & $2.65 \pm 0.19$ \\
 NRQM \cite{NRQM}& $2.41\pm 0.07\pm 0.02$& $2.79 \pm 0.08 \pm 0.02$& $2.37\pm 0.07\pm0.02$ \\
 \hline\hline
\end{tabular}
\end{center}
%\begin{center}
\caption{Decay widths $\Gamma(\Sigma_c \to \Lambda_c \pi)$ (in units of MeV). Experimental data and
other theoretical calculations are also shown.}
\end{table}
\begin{table}[htb]
\begin{center}
\begin{tabular}{c|c|c|c} \hline \hline
 & $\Gamma(\Sigma^{*++}_c\to\Lambda^+_c \pi^+)$ & $\Gamma(\Sigma^{*+}_c\to\Lambda^+_c \pi^0)$ &
 $\Gamma(\Sigma^{*0}_c\to\Lambda^+_c \pi^-)$
  \\ \hline
 experiment \cite{PDG06} & $14.9\pm 1.9$ & $< 17 (\text {CL} =90\%)$ & $16.1\pm 2.1$    \\
 \textbf{this work} & 14.7 & 15.2 & 14.5\\
 CQM  \cite{CQM1}& $20$ & $20$ & $20$ \\
 HHCPT \cite{HHCPT2}& 25 & 25 & 25 \\
 HHCPT \cite{HHCPT4}& $16.7 \pm 2.3$ & $17.4 \pm 2.3$ & $16.6\pm 2.2$\\
 LFQM \cite{LFQM}& 12.84 & & 12.40 \\
 RTQM \cite{RTQM}& $21.99 \pm 0.87$ & & $21.21 \pm 0.81$ \\
 NRQM \cite{NRQM}& $17.52\pm 0.74\pm 0.12$& $17.31 \pm 0.73 \pm 0.12$& $16.90\pm 0.71\pm0.12$ \\
 \hline\hline
\end{tabular}
\end{center}
%\begin{center}
\caption{Decay widths $\Gamma(\Sigma^*_c \to \Lambda_c \pi)$ (in units of MeV). Experimental data and
other theoretical calculations are also shown.}
\end{table}
We find that the results are all consistent with the experimental data. In addition, if the value of
$m_{u,d}$ varies from $0$ to $8$ MeV and $R$ varies from $0.8$ fm to $1.2$ fm, then the decay width,
for example, $\Gamma(\Sigma^{++}_c \to \Lambda_c^+ \pi^+)$ varies just from $1.90$ to $1.94$ MeV.
That is, these calculations are insensitive to values of $m_{u,d}$ and $R$. We also present the
experimental data and other theoretical calculations in the tables.

Finally, we use the same parameters to predict the relevant strong decay widths in bottom sector.
Because the neutral bottomed baryons $\Sigma^{0(*)}_b$ have not found yet, we resort to the following
supposition to obtain $M_{\Sigma^{(*)0}_b}$: the mass differences $\Delta M_b$ among bottomed baryons
$\Sigma^+_b$, $\Sigma^0_b$, and $\Sigma^-_b$ come from two parts, one part is the quark mass
difference $m_d-m_u$, and the other part is the electromagnetic Coulomb energy. Thus we may take the
quark replacement $b \to s$ and observe the mass differences among $\Sigma^+$, $\Sigma^0$, and
$\Sigma^-$. From Ref. \cite{PDG06}, it is easily to find that
 \be
 M_{\Sigma^0} \simeq \frac{1}{2} (M_{\Sigma^+}+M_{\Sigma^-}),
 ~~~M_{\Sigma^{*0}} \simeq \frac{1}{2} (M_{\Sigma^{*+}}+M_{\Sigma^{*-}}).
 \en
The errors are both smaller than $10^{-3}$. Therefore, we may approximately use the similar equations
 \be
 M_{\Sigma^0_b} \to \frac{1}{2} (M_{\Sigma^+_b}+M_{\Sigma^-_b}),
 ~~~M_{\Sigma^{*0}_b} \to \frac{1}{2} (M_{\Sigma^{*+}_b}+M_{\Sigma^{*-}_b}).
 \en
The results of the decay widths $\Gamma(\Sigma_b^{(*)} \to \Lambda_b \pi)$ are listed in TABLE IV.

\begin{table}[htb]
\begin{center}
\begin{tabular}{c|c|c||c|c|c} \hline \hline
 $\Gamma(\Sigma^{+}_b\to\Lambda^0_b \pi^+)$ & $\Gamma(\Sigma^{0}_b\to\Lambda^0_b \pi^0)$ &
 $\Gamma(\Sigma^{-}_b\to\Lambda^0_b \pi^-)$ & $\Gamma(\Sigma^{*+}_b\to\Lambda^0_b \pi^+)$ & $\Gamma(\Sigma^{*0}_b\to\Lambda^0_b \pi^0)$ &
 $\Gamma(\Sigma^{*-}_b\to\Lambda^0_b \pi^-)$
  \\ \hline
 $4.35$ & $5.65$ & $5.77$ & 8.50 & 10.20 & 10.44\\ \hline \hline
\end{tabular}
\end{center}
%\begin{center}
\caption{The predictions of the strong decay widths $\Gamma(\Sigma^{(*)}_b \to \Lambda_b \pi)$ (in
units of MeV).}
\end{table}

%%%%%%%%%%%%%%%%%%%%%%%%%
\section{Conclusion}
In this paper we have presented a formalism to describe the chiral dynamics of the heavy baryons
interacting with the Goldstone bosons. Furthermore, through PCAC, we obtained the relevant strong
constants in the MIT bag model, and then estimated the strong decay widths $\Gamma(\Sigma^{(*)}_{b,c}
\to \Lambda_{b,c} \pi)$. The parameters appearing in this approach are the light quark masses
$m_{u,d}$ and the bag radius $R$. On the one hand the current masses of $u,d$ quarks are both very
small on the typical hadron energy scale and can be taken as zero, and on the other hand the bag
radius $R=1$ fm was obtained by fitting the mass spectrum in the charm \cite{BS} and bottom
\cite{Ponce} sectors and averaging them. In addition, we also separately vary $m_{u,d}$ from 0 to 8
MeV and $R$ from 0.8 to 1.2 fm, and find the decay widths are insensitive to these two parameters.
Base on the numerical results being all consistent with the experimental data in charm sector, we
make the predictions on the strong decay widths $\Gamma(\Sigma^{(*)}_b \to \Lambda_b \pi)$. In spite
that the bottomed baryons $\Sigma_b^{(*)0}$ have not been found, we observed that the masses of
neutral baryons $M_{\Sigma^{(*)0}}$ almost equal to the average of the ones of charged baryons
$M_{\Sigma^{(*)+}}$ and $M_{\Sigma^{(*)-}}$, and then analogized to the masses of neutral bottomed
baryons $M_{\Sigma_b^{(*)0}}$. We expect the deviations of this assumption and the predictive strong
decay widths are all small for the future experimental data.

{\bf Acknowledgments}\\
 I would like to thank Dr. Chun-Khiang Chua for helpful discussion. This work
 is supported in part by the National Science Council of R.O.C. under Grant No:
 NSC-95-2112-M-017-001.

\end{document}